\documentclass[doublecol]{epl2}
\usepackage{color}
\usepackage{graphicx}
\usepackage{dcolumn}
\usepackage{bm}
\usepackage{array}
\usepackage[table]{xcolor}

\newcommand{\PreserveBackslash}[1]{\let\temp=\\#1\let\\=\temp}
\newcolumntype{C}[1]{>{\PreserveBackslash\centering}p{#1}}
\newcolumntype{R}[1]{>{\PreserveBackslash\raggedleft}p{#1}}
\newcolumntype{L}[1]{>{\PreserveBackslash\raggedright}p{#1}}
\begin{document}

\title{Relative clock demonstrates the endogenous heterogeneity of human dynamics}

\author{Tao Zhou\inst{1} \and Zhi-Dan Zhao\inst{1,2} \and Zimo Yang\inst{1,3}\and Changsong Zhou\inst{2}}
\shortauthor{Tao Zhou \etal} \institute{ \inst{1}Web Sciences
Center, University of Electronic Science and Technology of China,
Chengdu 610054, People's Republic of China\\
\inst{2}Department of Physics, Centre for Nonlinear Studies, and
Beijing-Hong Kong-Singapore Joint Centre for Nonlinear and Complex
Systems (Hong Kong), Hong Kong Baptist University, Kowloon Tong,
Hong Kong, People's Republic of China \\
\inst{3}Department of Physics, The Chinese University of Hong Kong,
Shatin, Hong Kong, People's Republic of China}


\abstract {The heavy-tailed inter-event time distributions are
widely observed in many human-activated systems, which may result
from both endogenous mechanisms like the highest-priority-first
protocol and exogenous factors like the varying global activity
versus time. To distinguish the effects on temporal statistics
from different mechanisms is this of theoretical significance. In
this Letter, we propose a new timing method by using a relative
clock, where the time length between two consecutive events of an
individual is counted as the number of other individuals' events
appeared during this interval. We propose a model, in which agents
act either in a constant rate or with a power-law inter-event time
distribution, and the global activity either keeps unchanged or
varies periodically versus time. Our analysis shows that the heavy
tails caused by the heterogeneity of global activity can be
eliminated by setting the relative clock, yet the heterogeneity
due to real individual behaviors still exists. We perform
extensive experiments on four large-scale systems, the search
engine by AOL, a social bookmarking system--\emph{Delicious}, a
short-message communication network, and a microblogging
system--\emph{Twitter}. Strong heterogeneity and clear seasonality
of global activity are observed, but the heavy tails cannot be
eliminated by using the relative clock. Our results suggest the
existence of endogenous heterogeneity of human dynamics.}

\maketitle

\section{Introduction}

Characterizing and understanding human activity patterns are
necessary to explain many socioeconomic phenomena and could find
significant applications ranging from resource allocation to
transportation control, from epidemic prediction to interface
design for Internet users \cite{Barabasi2007,Zhou2008}. One of the
most attractive observations is the heavy-tailed nature of human
temporal activities, with the inter-event time distribution
usually being approximate to a power-law form. Example include the
email communication \cite{Barabasi2005}, the surface mail
communication \cite{Oliveira2005,Li2008}, the cell-phone
communication \cite{Candia2008,Hong2009,Zhao2011}, the online
activities \cite{Dezso2006,Zhou2008b,Goncalves2008,Radicchi2009},
and so on, to name just a few.

Many endogenous mechanisms of human activities have been put
forward to explain the observed heavy-tailed statistics, such as
the task priority \cite{Barabasi2005,Vazquez2006}, the varying
interest \cite{Han2008,Shang2010}, the memory effects
\cite{Vazquez2007}, the human interacting
\cite{Oliveira2009,Min2009,Wu2010}, and so on. Besides the efforts
on digging out endogenous mechanisms, a litter pessimistic
argument is that the observed heavy-tailed statistics are hardly
to reveal significant ingredients or provide insights on human
activity patterns yet may originate from some trivial exogenous
factors\footnote{Here we use the word ``exogenous" to stand for
the factors not related to the essential motivations or
stimulations from the actions or other people.}. In particular,
the heterogeneity and seasonality\footnote{Denote by $M(T)$ the
global activity of the population (i.e., the number of events
during the $T$'s time window), the heterogeneity lies in the
heterogeneity of the distribution of $M$, and the seasonality is
evidenced by $M(T)\approx M(T+P)$, where $P$ is the time period,
normally being a day and/or a week in our daily behaviors.} of
human activities has recently been recognized as one candidate to
explain the heavy-tailed inter-event time distribution
\cite{Hidalgo2006,Malmgren2008}. Putting the mathematics behind,
the basic idea is simple: Take the short-message communication as
an example, an individual usually does not send messages during
the sleeping time, which forms large time intervals, and compared
with frequent communications in the day time, these intervals
spanning across the midnight contribute to the heavy tails.
Accordingly, the observed statistical regularities may result from
hybrid mechanisms \cite{Kentsis2006}--some of them are endogenous
like the highest-priority-first rule \cite{Barabasi2005}, while
others are exogenous like activity heterogeneity
\cite{Hidalgo2006} and seasonality \cite{Malmgren2008}.

In this Letter, we propose a new timing method that can eliminate
the heavy tails in the inter-event time distribution caused by the
activity heterogeneity. We analyze a model, in which agents act
with either an exponential or a power-law inter-event time
distribution, and the global activity either keeps unchanged or
varies periodically versus time. Simulation results show that the
heavy tails caused by the heterogeneity of activity can be
eliminated by setting the relative clock, yet the heterogeneity
due to endogenous individual behaviors still exists. Comparing the
modeling results to the experiments on four large-scale real
systems, we conclude that the temporal activity contains
endogenous heterogeneity that cannot be explained by Poissonian
agent assumption with seasonality.

\begin{figure}
\begin{center}
\includegraphics[width=8.5cm]{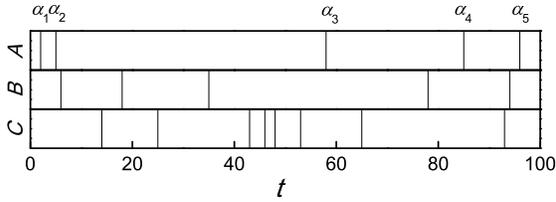}
\caption{An illustration about the absolute clock and relative
clock for the definitions of inter-event time. $A$, $B$ and $C$
refer to three different individuals and the vertical lines stand
for actions.}
\end{center}
\end{figure}

\begin{table}
\caption {Inter-event times for individual $A$ in figure 1. The
upper row corresponds to the results based on the absolute clock
while the lower row on the relative clock. In the case of relative
clock, we use the number plus one to avoid zero interval.}
\centering
\begin{tabular}{ccccc}
\hline \hline
       & $(\alpha_1,\alpha_2)$ & $(\alpha_2,\alpha_3)$ & $(\alpha_3,\alpha_4)$ & $(\alpha_4,\alpha_5)$ \\
\hline
       Absolute  & 3 & 53 & 27 & 11\\
       Relative  & 1 & 10 & 3 & 3\\
\hline \hline
\end{tabular}
\end{table}

\begin{figure}
\begin{center}
\includegraphics[width=8.8cm]{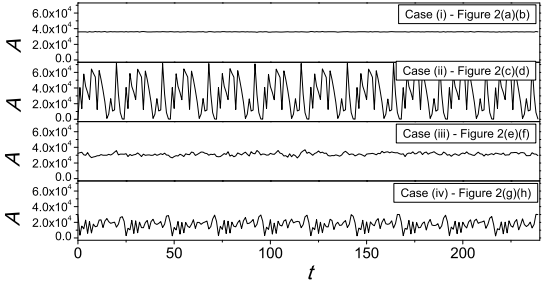}
\caption{How the global activity $A_i$, quantified by the number
of events happened during the $i$'s hour, changes with time. The
four plots respectively correspond to cases (i), (ii), (iii) and
(iv). The parameters are $N=100$, $r=0.2$, $\lambda=0.5$ and
$\beta=2$. Strong heterogeneity and clear daily seasonality are
observed for cases (ii) and (iv) yet only random fluctuations are
associated with cases (i) and (iii).}
\end{center}
\end{figure}

\section{Relative Clock}

The heterogeneity of human activity versus time has been observed
for many online systems. For example, we will later show four real
systems in Fig. 5. As we have mentioned above, the statistics
about inter-event time at the population level may result from
hybrid mechanisms, and thus it is valued to design a method that
can filter out the effects caused by the exogenous heterogeneity.
In the traditional way, the inter-event time is defined as the
time interval between two consecutive events. Figure 1 illustrates
a simple example where the individual $A$ acts at time
$\alpha_1=2$, $\alpha_2=5$, $\alpha_3=58$, $\alpha_4=85$ and
$\alpha_5=96$, and thus the four time intervals are
$\alpha_2-\alpha_1$, $\alpha_3-\alpha_2$, $\alpha_4-\alpha_3$ and
$\alpha_5-\alpha_4$. This timing method is called \emph{absolute
clock} in this Letter. Considering a system with strong
heterogeneity of human activity versus time. For example, in a
short-message communication network, an individual may send in
average more than ten messages in the noon yet less than one
message during the midnight. As a time interval, $1h$ is
relatively long in the noon yet $10h$ is usual across the
midnight. Therefore, the absolute clock is highly affected by the
activity heterogeneity and thus may fail to capture the endogenous
human activity patterns. Accordingly, we propose a new timing
method by using a \emph{relative clock}, where the time length
between two consecutive events of an individual is counted as the
number of other individuals' events appeared during this interval.
Considering the population $A$, $B$ and $C$ shown in Fig. 1, the
inter-event time of the events happened at $\alpha_2$ and
$\alpha_3$ for individual $A$ is counted as the number of events
in between $\alpha_2$ and $\alpha_3$ for individuals $B$ and $C$.
Table 1 presents the results of two definitions of the inter-event
time for individual $A$. Compared with the absolute clock, the
relative clock, running faster at the time with frequent events,
can be considered as a kind of time rescaling method that can
eliminate the heavy tails of inter-event time distribution caused
by the activity heterogeneity.

\begin{figure}
\begin{center}
\includegraphics[width=4.2cm]{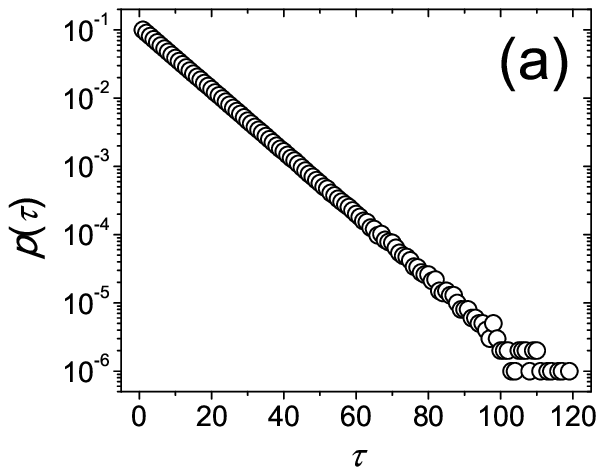}
\includegraphics[width=4.2cm]{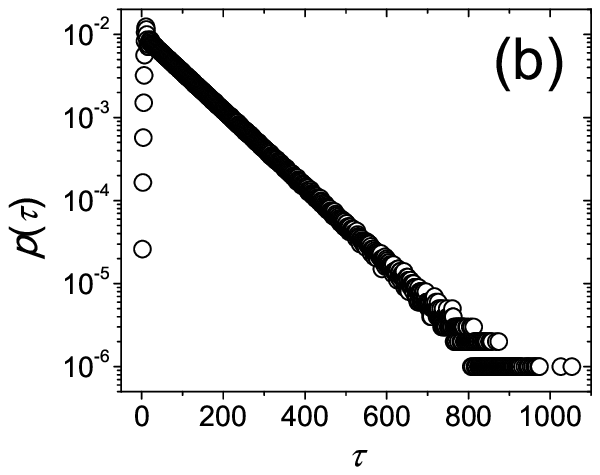}
\includegraphics[width=4.2cm]{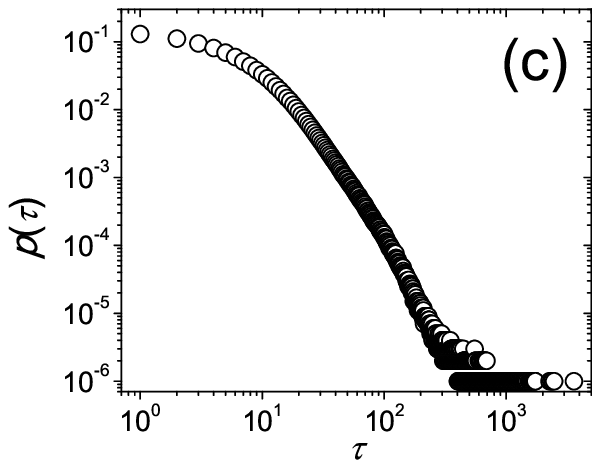}
\includegraphics[width=4.2cm]{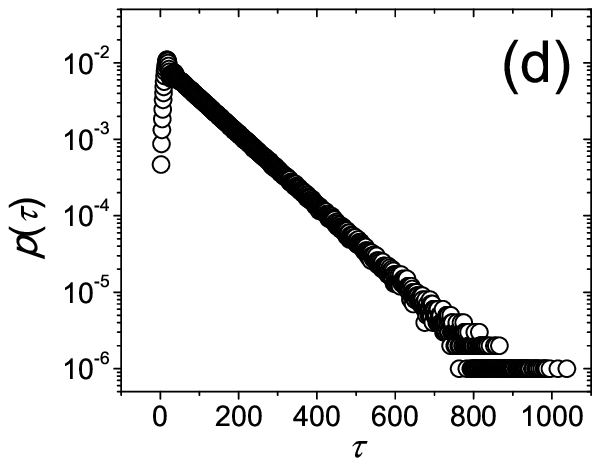}
\includegraphics[width=4.2cm]{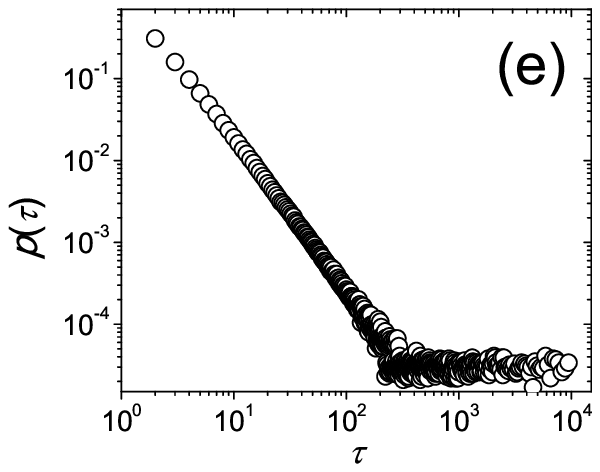}
\includegraphics[width=4.2cm]{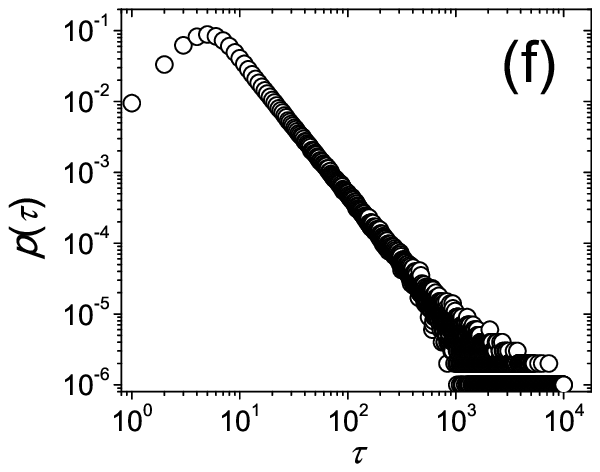}
\includegraphics[width=4.2cm]{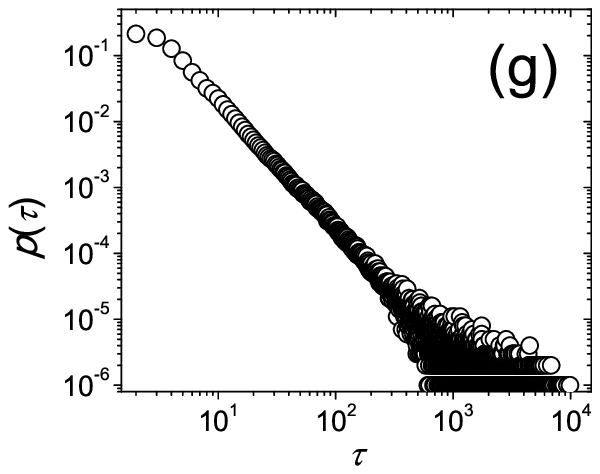}
\includegraphics[width=4.2cm]{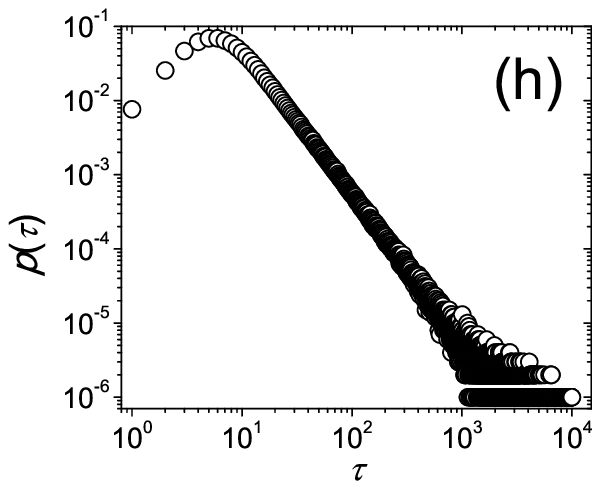}
\caption{Comparison of inter-event time distributions $p(\tau)$
based on the absolute and relative clocks. All the distributions
presented in this figure come from the theoretical model: case
(i)--(a)(b), case (ii)--(c)(d), case (iii)--(e)(f), case
(iv)--(g)(h). The left and right plots correspond to the
distributions on absolute and relative clocks, respectively. Plots
(a), (b) and (d) are of log-linear scale, while plots (c), (e),
(f), (g) and (h) are of log-log scale. The parameters are $N=100$,
$r=0.2$, $\lambda=0.5$ and $\beta=2$. The power-law sampling on
$\Phi(t)$ follows the method in Ref. \cite{Clauset2009}.}
\end{center}
\end{figure}

\section{Model}

To see the difference between absolute and relative clocks, we
first study a theoretical model. This model spans over 10 days,
with a second resolution, namely it contains 864000 time steps.
Each day is divided into 24 hours, and for simplicity, the global
activity inside an hour keeps unchanged. Accordingly, for each
hour $i$, we denote its activity as $\lambda_i$. For the first
day, the value of $\lambda$ for each of the 24 hours is sampled
from a given distribution $\Psi(\lambda)$. To account for the
seasonality, the following 9 days will repeat the activity pattern
of the first day, that is, $\lambda_i=\lambda_{i+24}$. All the $N$
individuals in the model have the same temporal statistics. We
consider four cases: (i) Every individual follows a Poissonian
process with rate $r$, that is, at each second, an arbitrary
individual $A$ will act with probability $r\lambda_i$, where $i$
denotes the current hour. We assume a constant global activity,
say $\lambda_1=\lambda_2=\cdots=\lambda_{24}=\lambda$. (ii) Same
to the case (i), but $\lambda_i$ $(i=1,2,\cdots,24)$ are
independently sampled from a uniform distribution in the range
$(0,1)$, say $\Psi(\lambda)=U(0,1)$. (iii) Every individual acts
with an endogenous power-law inter-event time distribution
$\Phi(t)\sim t^{-\beta}$. In the beginning, each individual will
sample an inter-event time $t$ from $\Phi(t)$, and will indeed act
at time $t/\lambda_1$. Then, after each act, the individual will
resample an inter-event time $t'$ from $\Phi(t)$ and act after
$t'/\lambda_i$ seconds, where $i$ denotes the current hour. This
rule reflects the fact that in an inactive time period, the
inter-event time tends to be longer, and vice verse. If the time
interval spans over more than one hour, only the activity of the
starting hour affects the real length of the time interval. We
assume $\lambda_1=\lambda_2=\cdots=\lambda_{24}=\lambda$. (iv)
Same to the case (iii), but $\lambda_i$ $(i=1,2,\cdots,24)$ are
independently sampled from the uniform distribution $U(0,1)$.

\begin{figure}
\begin{center}
\includegraphics[width=4.2cm]{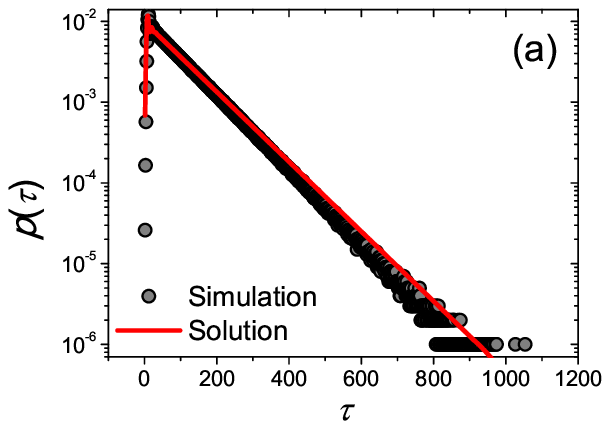}
\includegraphics[width=4.2cm]{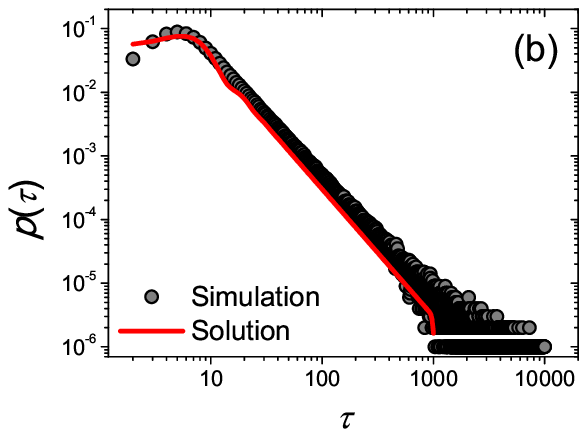}
\caption{(Color online) Analytical results about the inter-event
time distributions on relative clock. The plots (a) and (b)
correspond to Fig. 2(b) and 2(f) respectively, with black circles
representing the simulation results and red curves standing for
the analytical solutions.}
\end{center}
\end{figure}

Figure 2 displays the global activity $A_i$ $(i=1,2,\cdots,240)$
for the 240 hours, where $A_i$ is the number of total events in
the $i$'s hour. For the cases (i) and (iii), the global activity
is homogeneous, and thus the relative clock will not change the
overall statistical regularities although it can to some extent
reduce the fluctuation. The heterogeneity of global activity is a
necessary condition for the elimination of the heavy tail by using
the relative clock, yet not a sufficient condition.

\begin{table}
\caption {The number of users and the number of events in the four
real data sets. The last column gives the original places of the
used data sets, with the last two data sets are firstly reported
in this Letter.} \rowcolors{2}{gray!20}{white} \centering
\begin{tabular}{cccc}
\hline \hline
       Data Sets & \#Users & \#Events & Origins \\
\hline
       AOL  & 356610 & 4596212 & \cite{Radicchi2009}\\
       Delicious  &  256676 & 1252947 & \cite{Lu2011} \\
       SM & 1479480 & 28951117 & This Letter\\
       Twitter & 2711178 & 9966800 & This Letter\\
\hline \hline
\end{tabular}
\end{table}

Figure 3 reports the simulation results for the toy model, from
which we conclude that: (i) As shown in Fig. 3(c), a
power-law-like inter-event time distribution could result from the
heterogeneity of global activity\footnote{We introduce a
periodical global activity to the model to mimic the seasonality
observed in the real systems
\cite{Zhou2008b,Radicchi2009,Malmgren2008}. However, the
seasonality does not essentially contribute to the heavy tail in
the inter-event time distribution. For example, if we assume
$\lambda_i$ $(i=1,2,\cdots,240)$ are independently sampled from
$U(0,1)$, then the seasonality will be eliminated yet the heavy
tail in the distribution $p(\tau)$ still exists.} although all
individuals are the same and each individual obeys a Poissonian
process in each hour. This is supportive to the theoretical
analyses of Refs. \cite{Hidalgo2006,Malmgren2008}. In fact,
endogenous factor, exogenous factors and the hybrid of them can
generate heavy tails in $p(\tau)$, as shown in Fig. 3(d), 3(f) and
3(h). (ii) As shown in Fig. 3(d), the inter-event time
distribution based on the relative clock follows an exponential
form, that is to say, the heavy-tail resulted from the
heterogeneity of global activity can be effectively eliminated by
using the proposed timing method. (iii) Comparing Fig. 3(f) with
3(e), as well as 3(h) with 3(g), it is clear that the endogenous
heterogeneity, embodied in the power-law distribution $\Phi(t)\sim
t^{-\beta}$, could not be eliminated by the timing with relative
clock. (iv) A peak near the head of $p(\tau)$ will emerge in all
the cases when using the relative clock.

\begin{figure}
\begin{center}
\includegraphics[width=4.2cm]{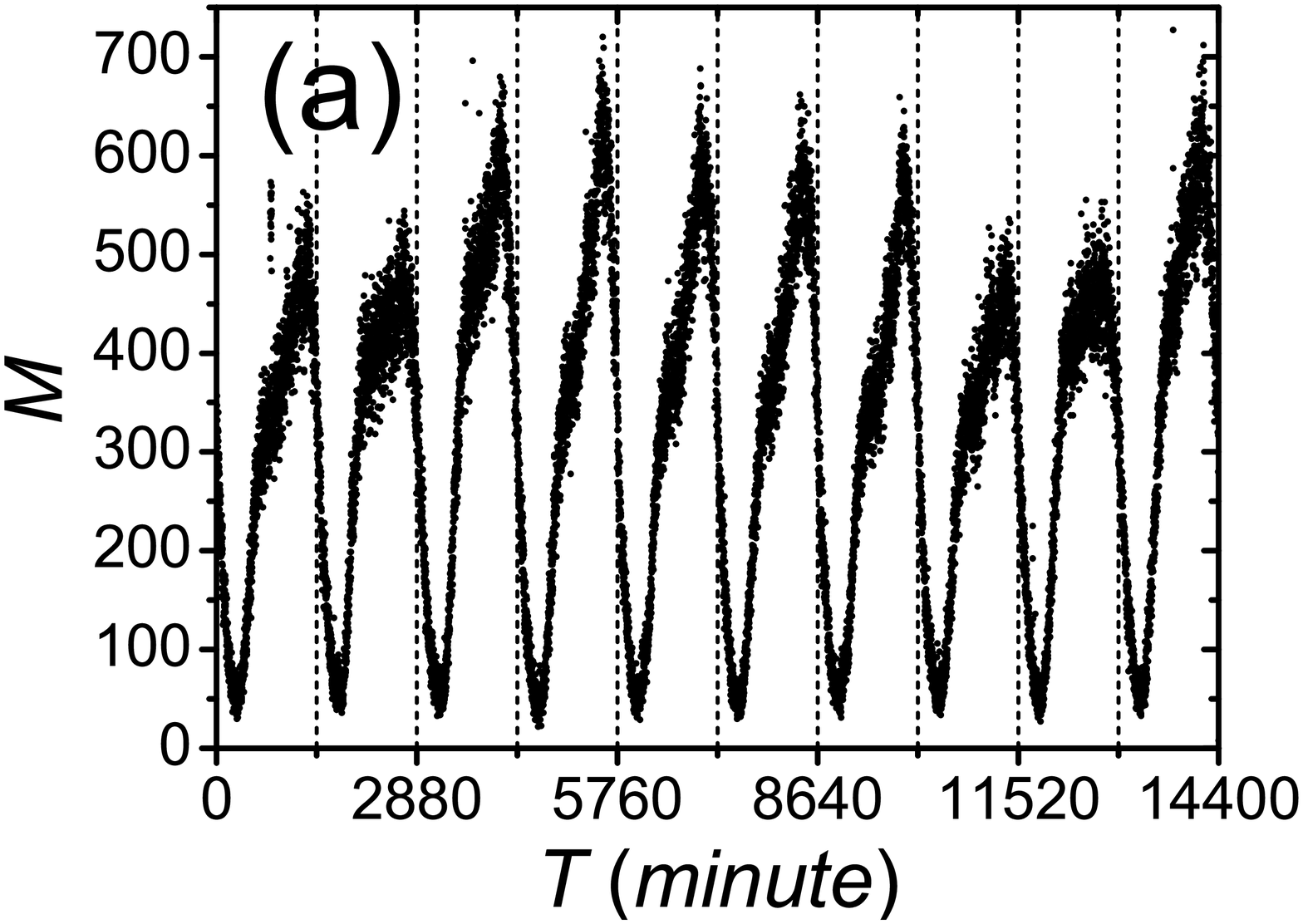}
\includegraphics[width=4.2cm]{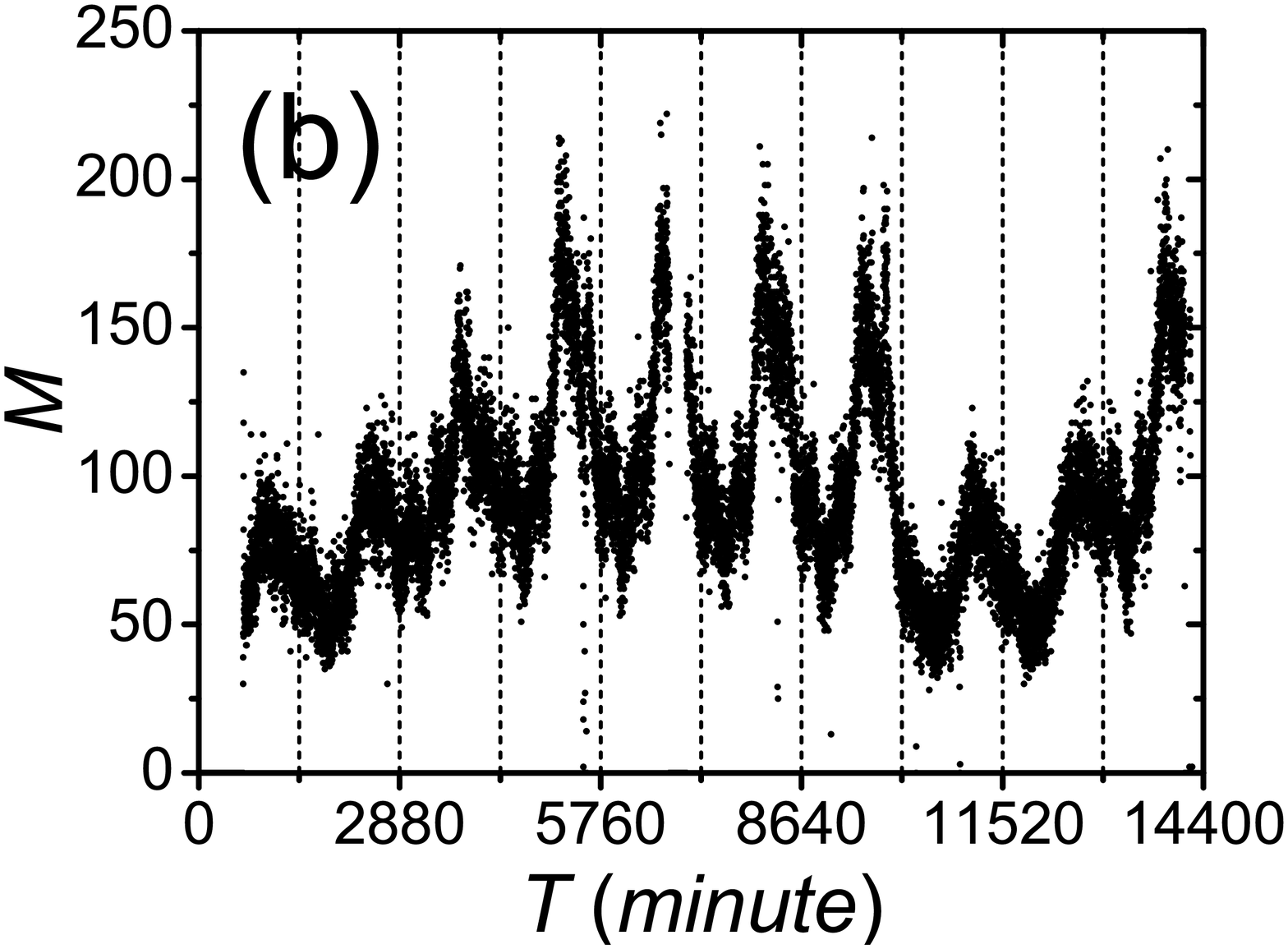}
\includegraphics[width=4.2cm]{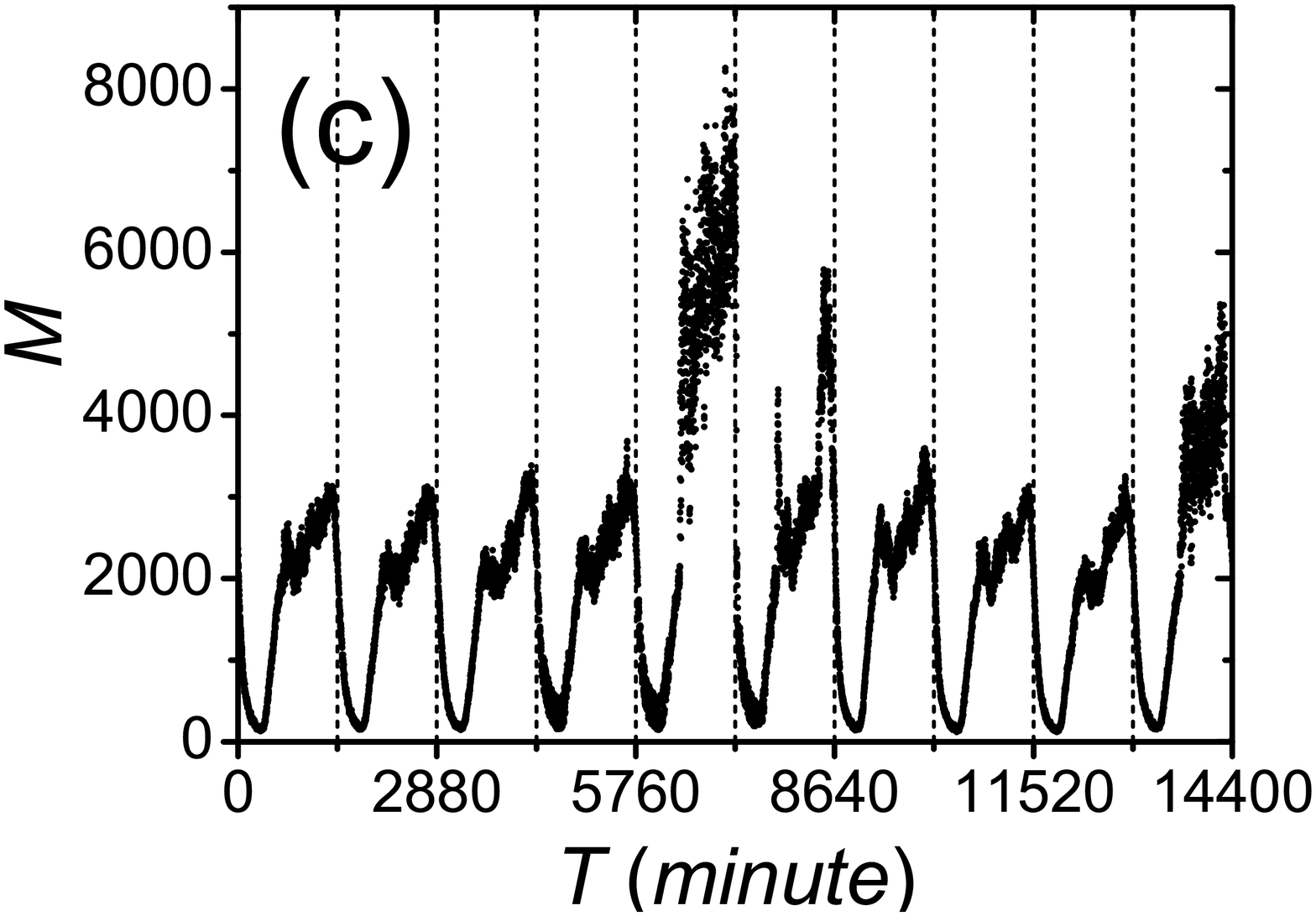}
\includegraphics[width=4.2cm]{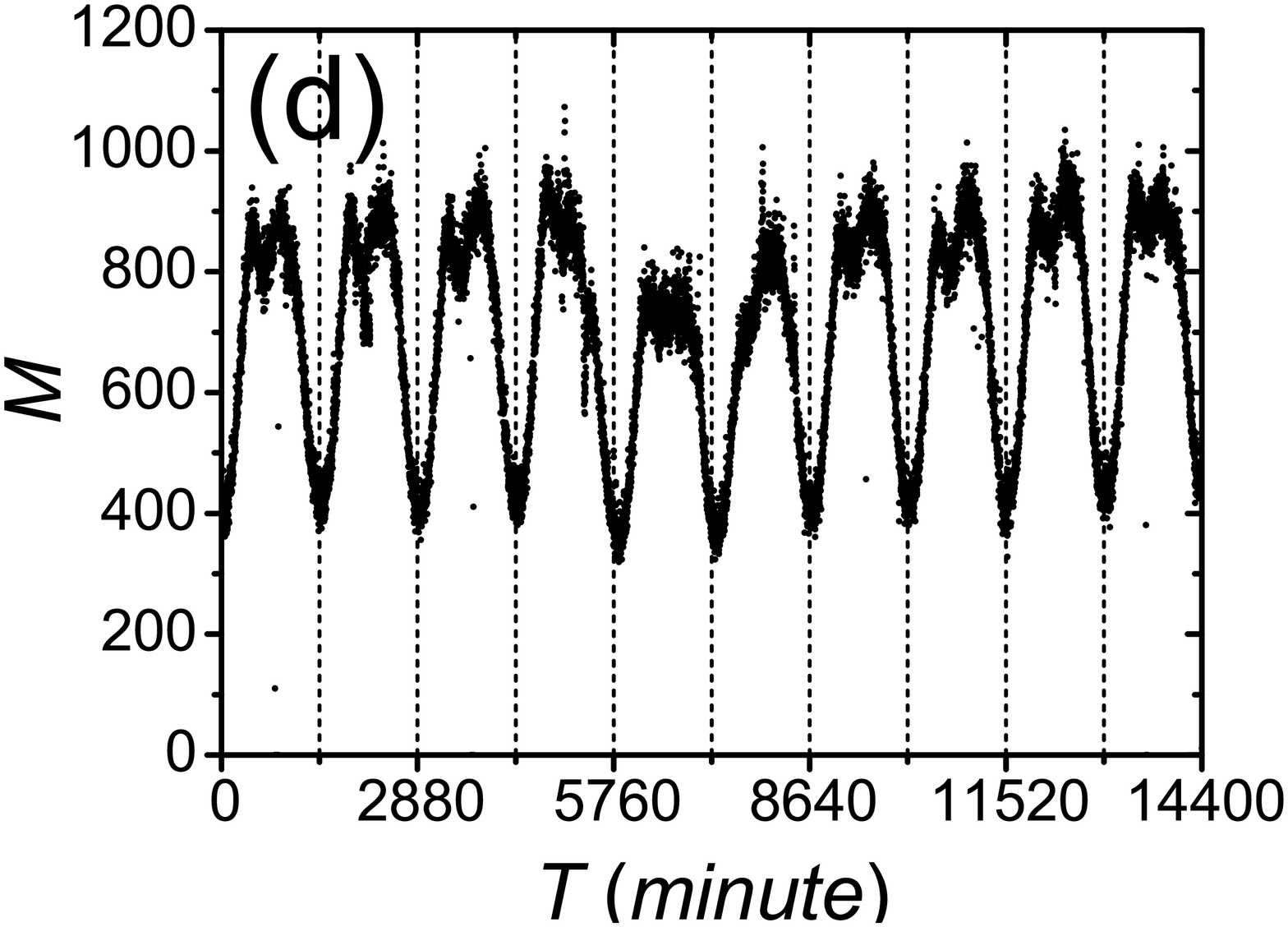}
\caption{The activity $M$ versus time with minute resolution for
AOL (a), Delicious (b), SM (c) and Twitter (d). The vertical dash
lines separate 10 days.}
\end{center}
\end{figure}

To explain the existence of a peak, we calculate the inter-event
time distribution $p(\tau)$ on relative clock. Notice that, since
the relative clock could eliminate the heterogeneity of global
activity, the idiographic form of $\Psi(\lambda)$ almost has
nothing to do with $p(\tau)$ (as an evidence, the distributions
shown in Fig. 3(b) and Fig. 3(d) are almost the same, and the
distributions shown in Fig. 3(f) and Fig. 3(h) are almost the
same). Considering two independent stochastic processes, the
actions of an individual and the actions of all others. Given a
monitored individual $i$, we assuming that her acting frequency
(i.e., the number of events during a unit time) is $f_i$, and the
total acting frequency of other individuals is
$f_{\bar{i}}=\sum_{j\neq i}f_j$, then the probability density of
the inter-event time of individual $i$ is:
\begin{equation}
p(t)=f_{i}e^{-f_{i}x}.
\end{equation}
Notice that, here we assume the individual $i$ at most act once in
one time step, namely $f_i<1$ and in each time step $i$ will
activate an event with probability $f_i$. In principle, we can
assume the time resolution is elaborate enough and thus at each
time step there is at most one event from all other individuals,
and the happening probability is $f_{\bar{i}}$. During $t$ time
steps, the probability density of the cumulative number of events
of all other individuals reads
\begin{equation}
q(a)=C_{t}^{a}f_{\bar{i}}^{a}(1-f_{\bar{i}})^{t-a},
\end{equation}
where $C_{t}^{a}=\frac{t!}{a!(t-a)!}$. When the activity of individual can be
approximated as a Poisson process, we can get the probability
distribution of the inter-event time on relative clock through the
joint probability distribution:
\begin{equation}
p(\tau)=\int_{0}^{\infty}
f_{\overline{i}}\frac{(f_{\overline{i}}-t)^{\tau -1}}{(\tau -1)!}e^{-f_{\overline{i}}t}e^{-f_{i}t}dt.
\end{equation}
Even when $f_{\overline{i}}>1$, We have checked numerically that
Eq. (3) can well reproduced the front peak in $p(\tau)$. For
example, in the case shown in Fig. 4(a), $f_{i}=0.1$ and
$f_{\overline{i}}=9.9$. Similar to the Poissonian cases, when the
endogenous time interval follows a power-law distribution, the
probability distribution of the inter-event time on relative clock
is:
\begin{equation}
p(\tau)=\int_{0}^{\infty}
f_{\overline{i}}\frac{(f_{\overline{i}}-t)^{\tau -1}}{(\tau -1)!}e^{-f_{\overline{i}}t}t^{-\beta}dt.
\end{equation}
where $\beta$ is the exponential power. Figure 4 reports the
analytical solutions Eq. (3) and Eq. (4), which agree very well
with the simulations.

\section{Data}

This Letter analyzes four large-scale real systems, and for fair
comparison, every data set presented here spans over 10 days.
Followed please find the data description, with basic statistics
shown in Table 2. (i) \emph{AOL}.-- It is previously known as
America Online, which is a company providing Internet services and
media, etc. This data set is about the searching behaviors of
Internet users, with time resolution being second. The date starts
from March 10, 2006 to March 20, 2006. The inter-event time is
defined as the time interval between two consecutive queries by a
user. (ii) \emph{Delicious}.-- It is a web site aiming at helping
users in collecting the tastiest bookmarks in the web. The data
set contains the bookmarks add by users with seconds resolution,
starting from September 5, 2009, last for 10 days. Each record
(i.e., event) contains the operation time, the users ID, the
Universal Resource Locator (URL), and so on. The inter-event time
is defined as the time interval between two consecutive
collections of bookmarks by a user. (iii) \emph{SM}.-- Short
Message is probably the most widely used electronic communication
tool in people's daily life. This data set starts from December
10, 2010 to December 20, 2010, with time resolution being second.
Each record consists of three elements: a sender ID, a receiver
ID, and the time stamp. The inter-event time is defined as the
time intervals between two consecutive short messages sent by the
same user. (iv) \emph{Twitter}.-- It is a microblogging system in
which users could upload their posts (i.e., microblogs) and other
users, especially their followers, may comment and/or transfer
these posts. The date starts from November 10, 2009, last for 10
days, with time resolution being second, recording only the
uploading time of original posts. The inter-event time is defined
as the time interval between two consecutive posts by a user.

\begin{figure}
\begin{center}
\includegraphics[width=4.2cm]{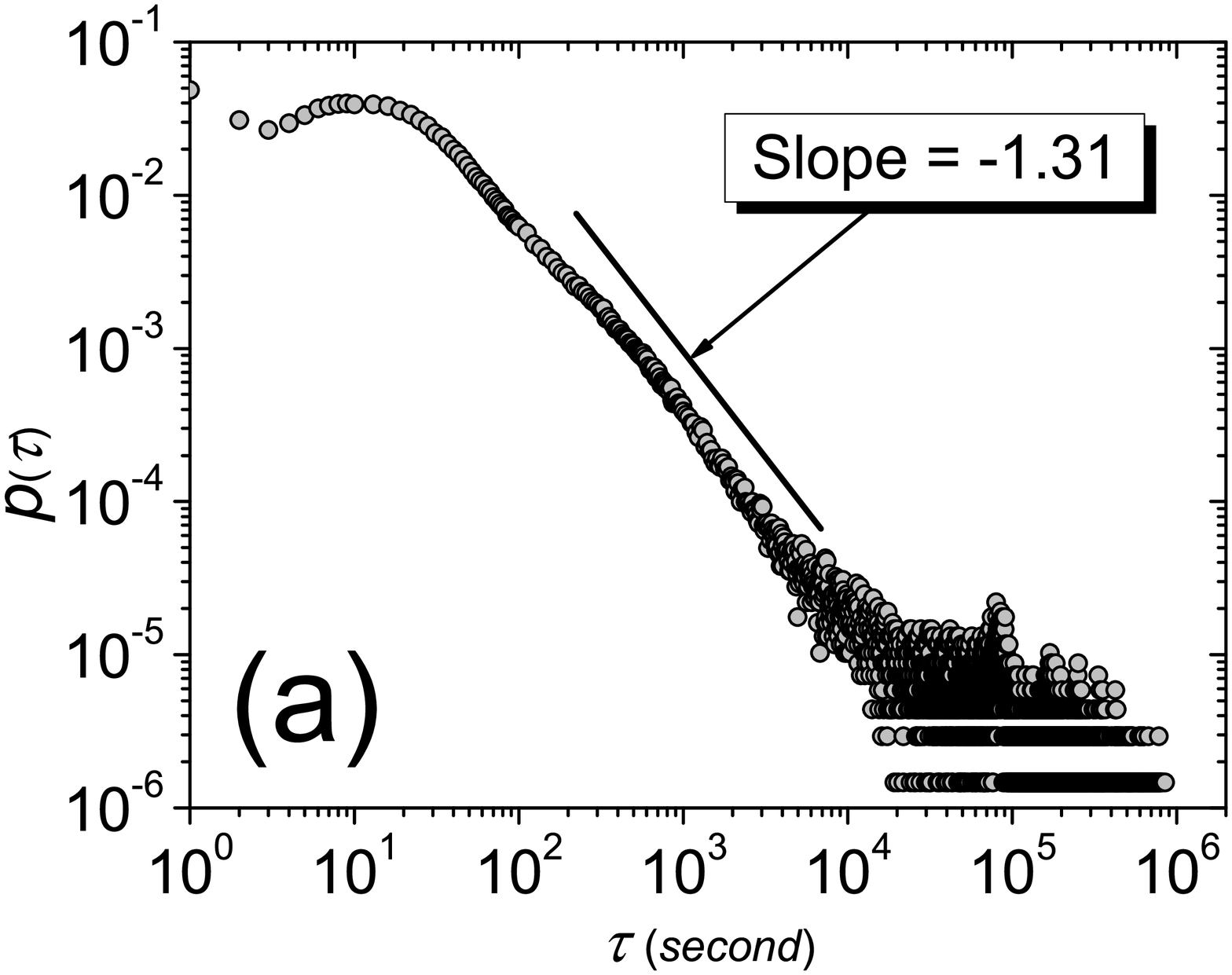}
\includegraphics[width=4.2cm]{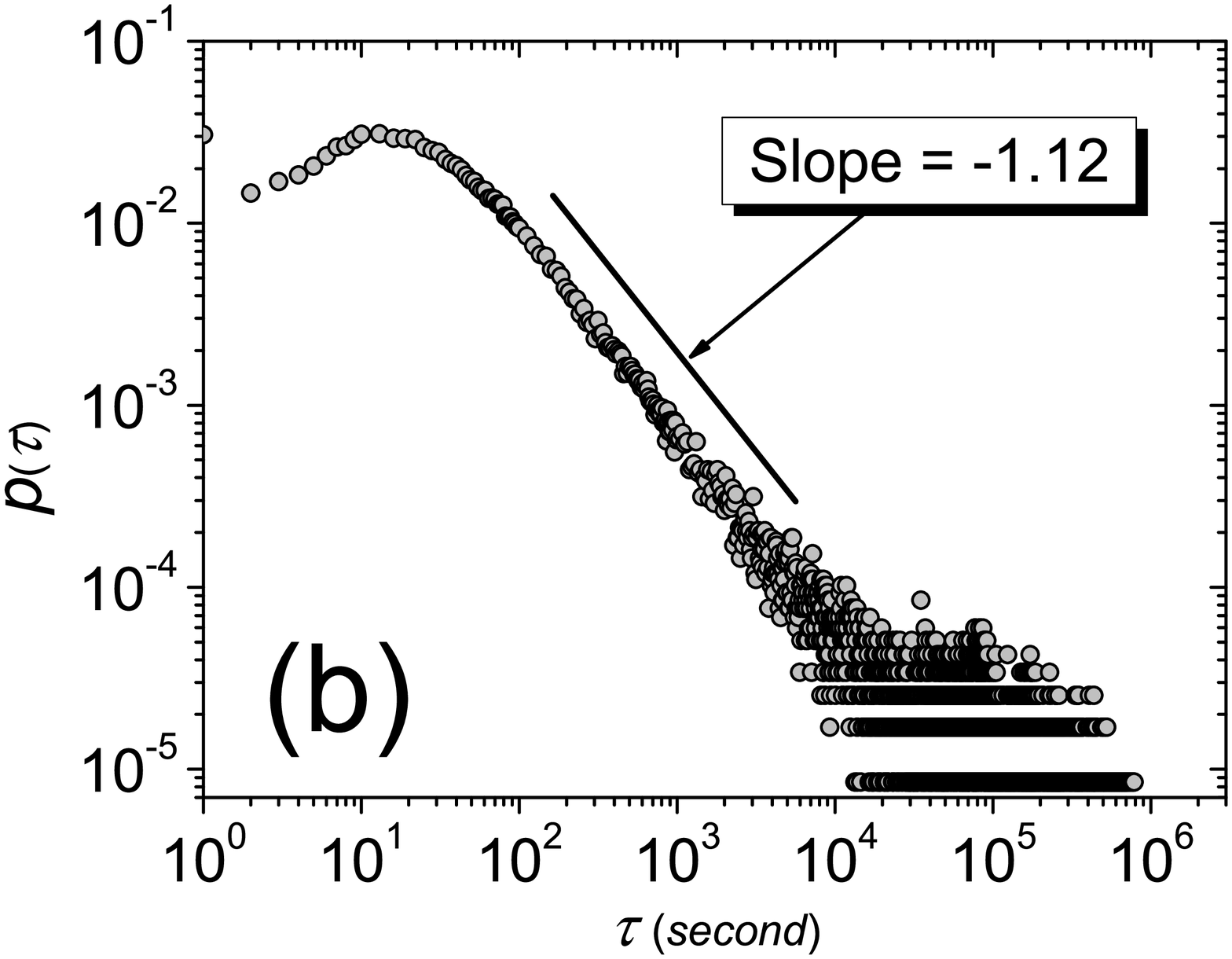}
\includegraphics[width=4.2cm]{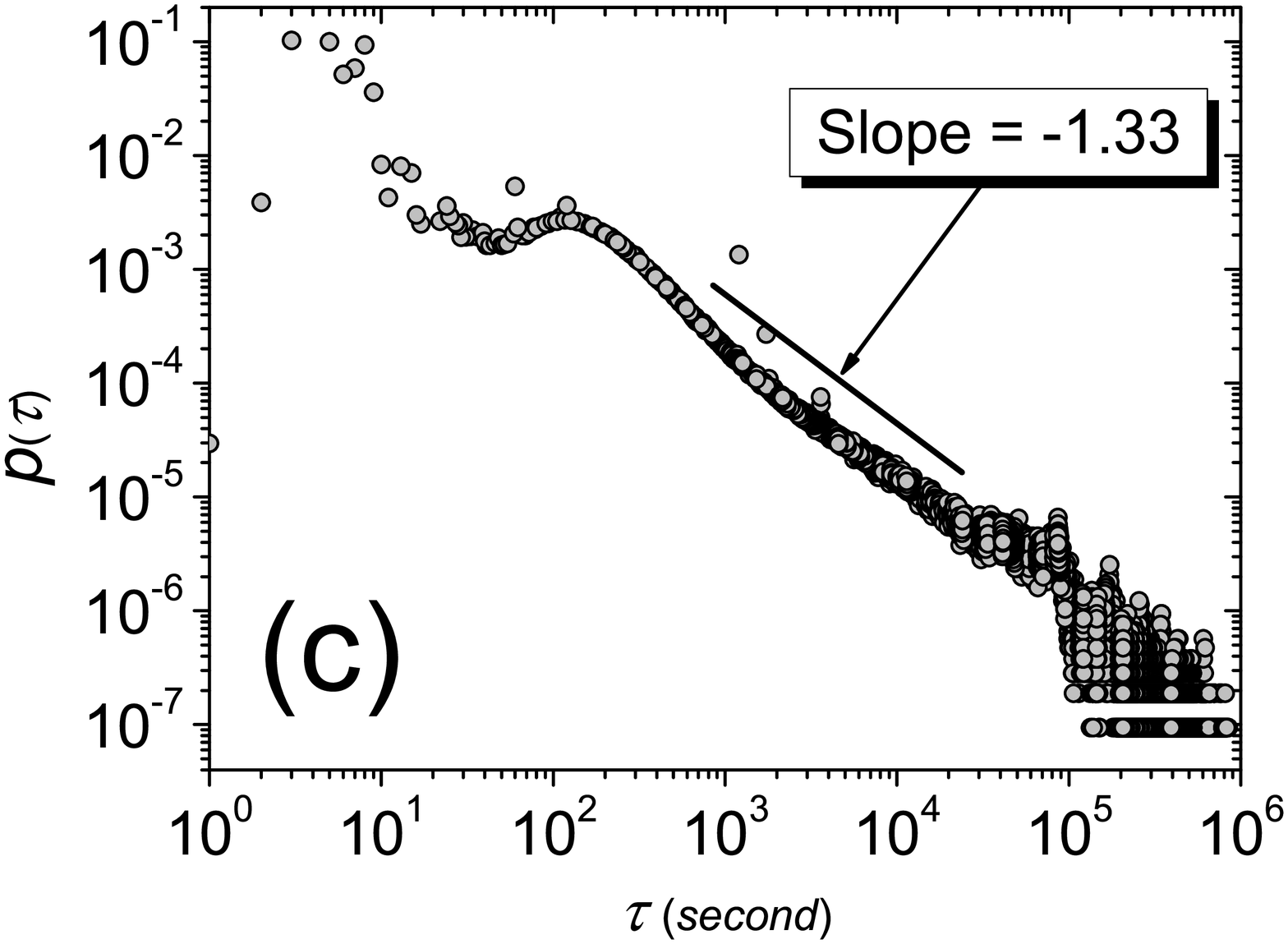}
\includegraphics[width=4.2cm]{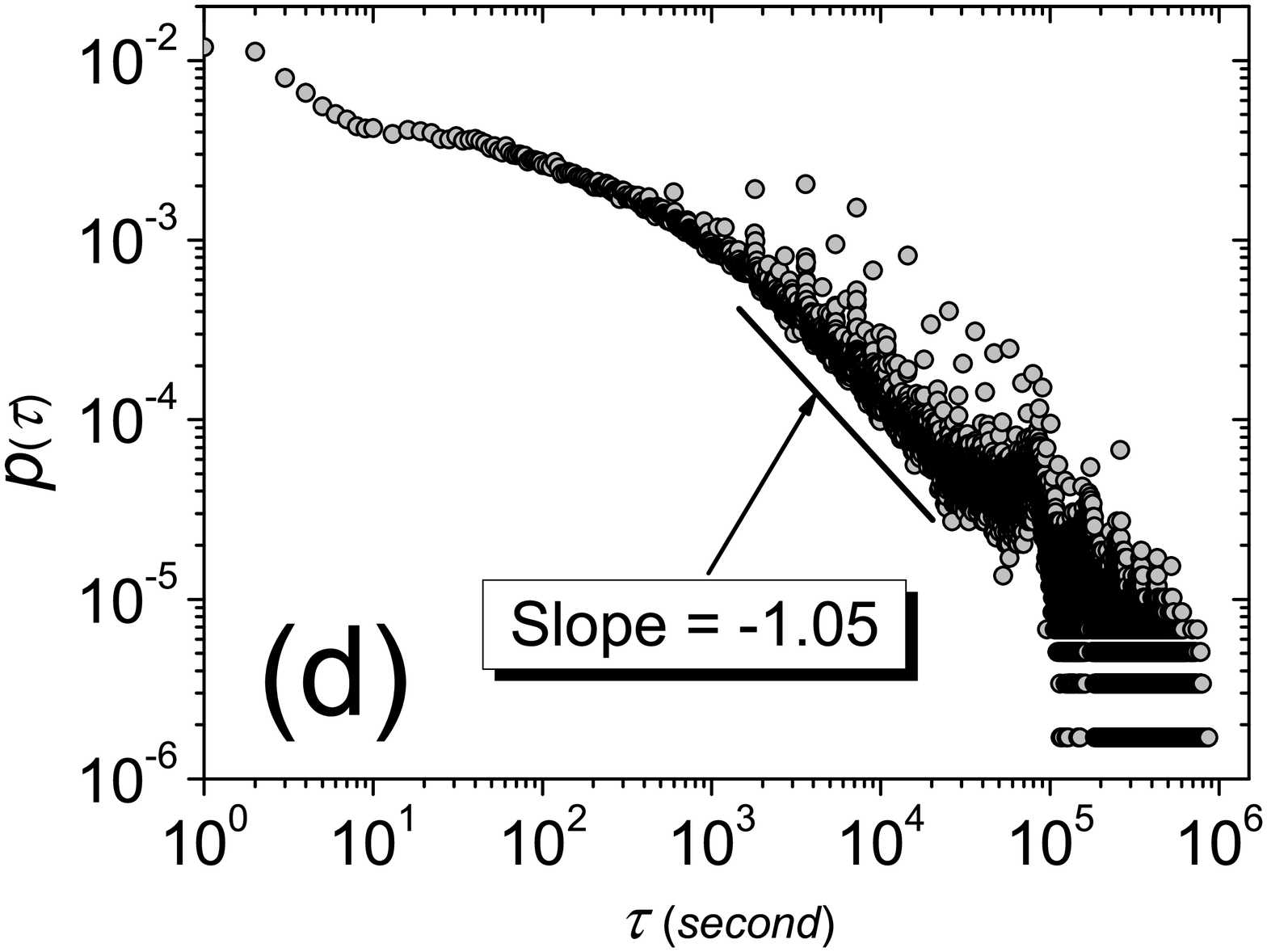}
\caption{Inter-event time distributions based on the absolute
clock for AOL (a), Delicious (b),  SM (c) and Twitter (d). These
curves partially display a power-law-like shape, yet they can not
be accurately fitted by simple power laws. The solid lines are
only for eye guidance.}
\end{center}
\end{figure}

\section{Experimental Results}

Figure 5 reports the global activity $M(T)$ versus the time $T$,
where the whole data is divided into 14400 segments, each of which
lasts one minute. That is to say, $M(T)$ is the number of event of
the population in $T$'s minute. It is observed that every system
displays strong heterogeneity\footnote{The typical difference of
the peaked and low-lying values of $M$ is about $10^2$ time in
AOL, Delicious and SM. This is really a huge. Even for Twitter,
the peaked value of $M$ can be as twice large as the low-lying
one.} and daily seasonality\footnote{Here we mainly concentrate on
the daily seasonality, yet for longer data, we could also observe
the weekly seasonality (see, e.g., the weekly seasonality in
Netflix \cite{Zhou2008b}).}.

\begin{figure}
\begin{center}
\includegraphics[width=4.2cm]{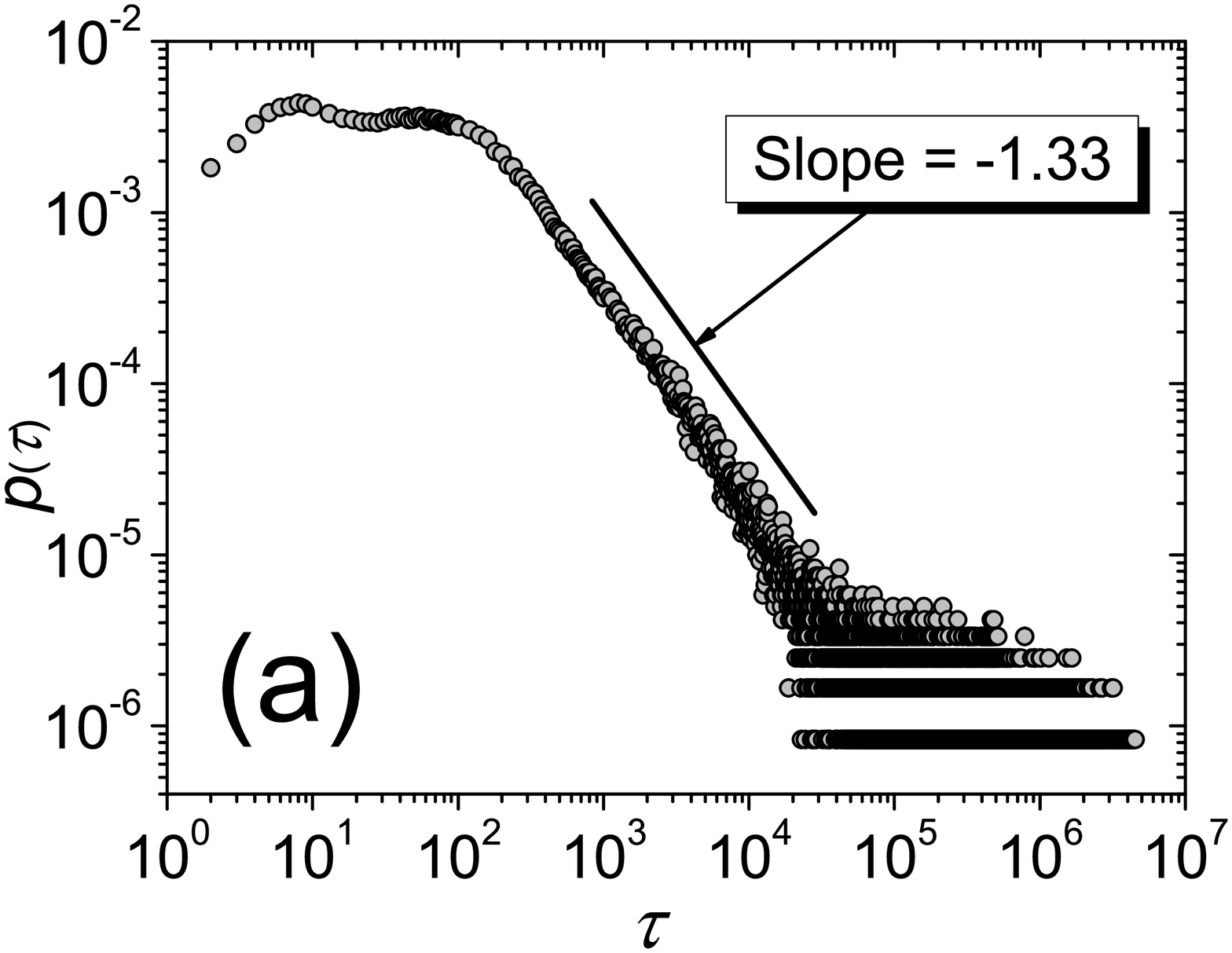}
\includegraphics[width=4.2cm]{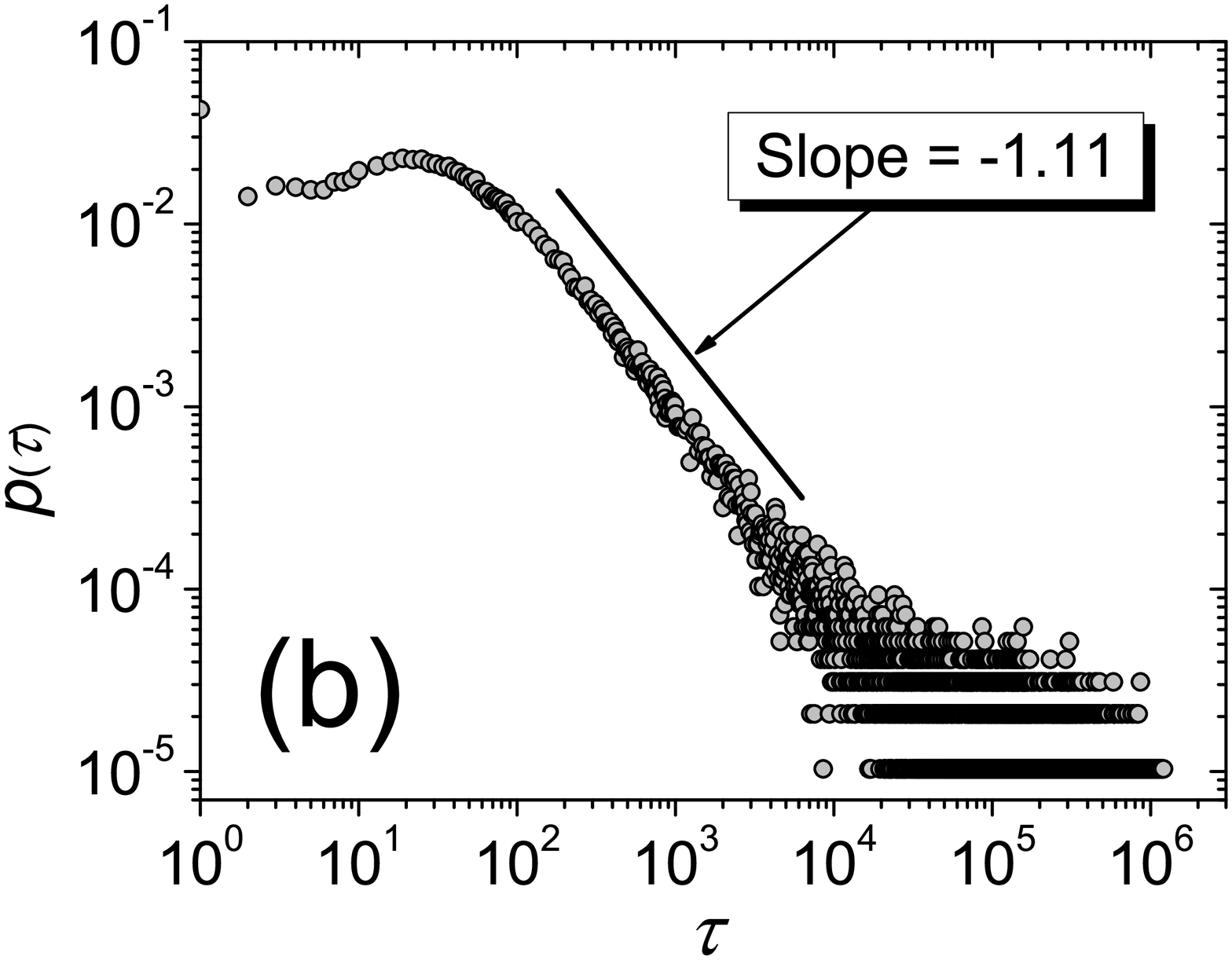}
\includegraphics[width=4.2cm]{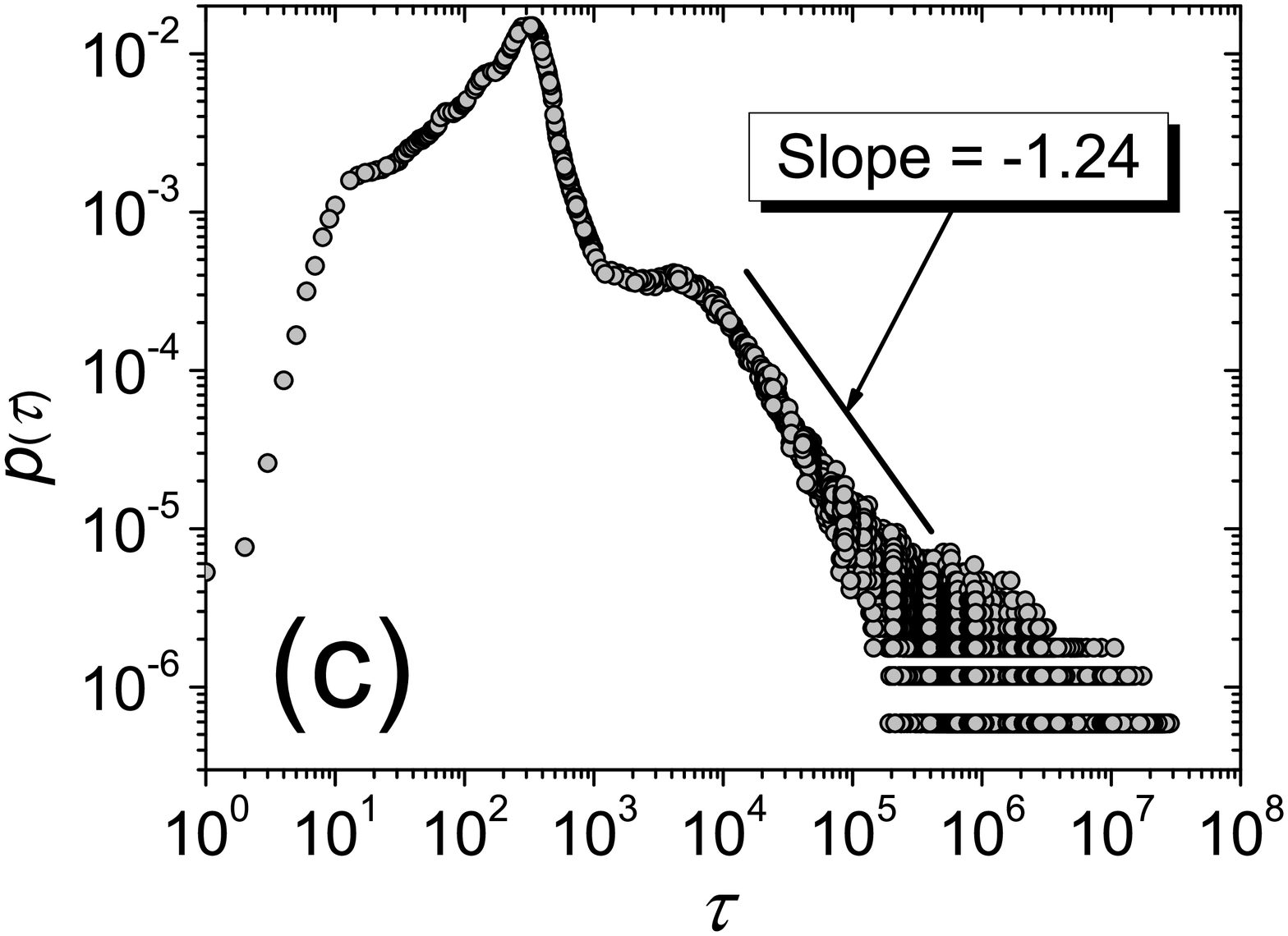}
\includegraphics[width=4.2cm]{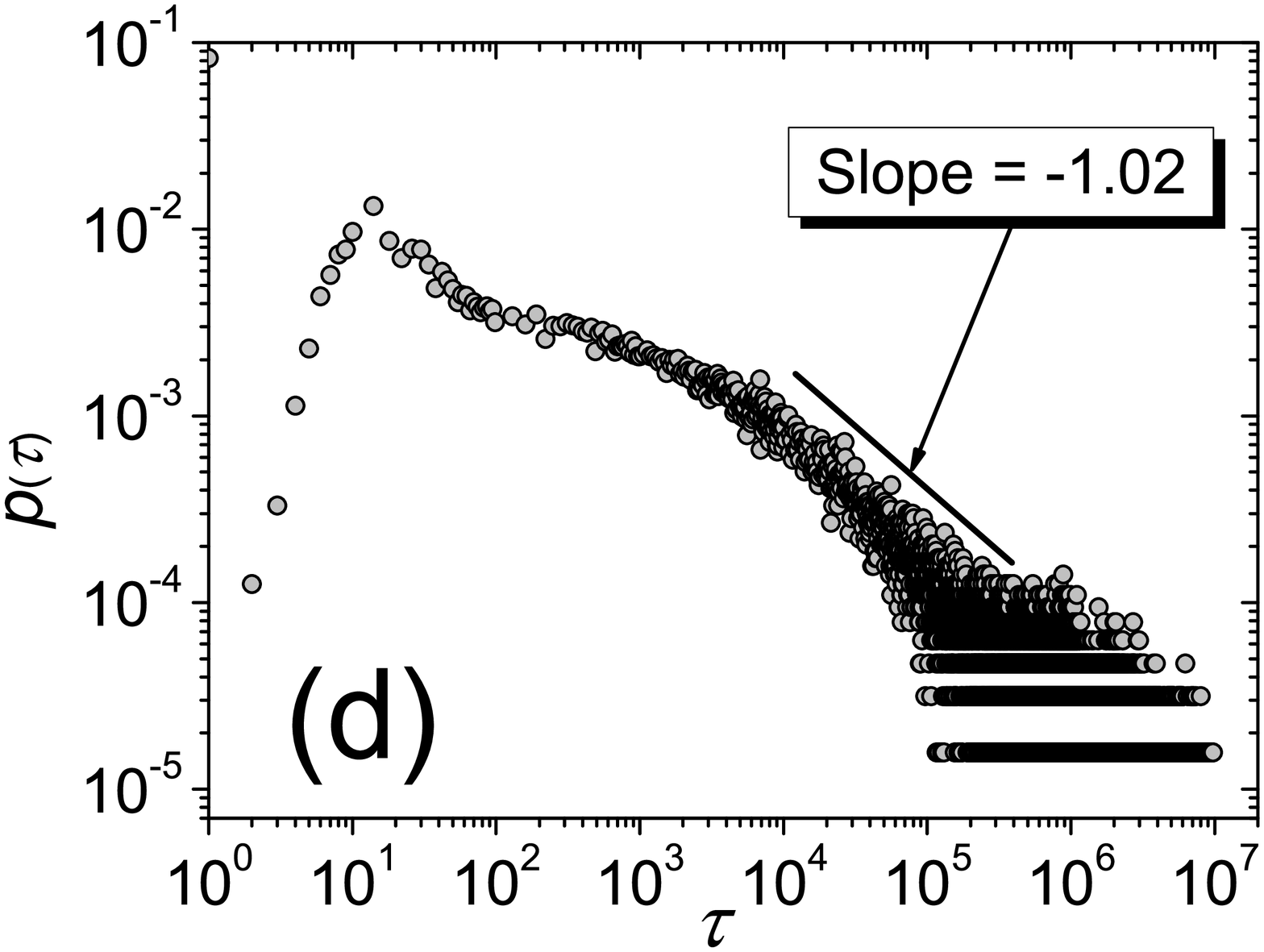}
\caption{Inter-event time distributions based on the relative clock
for AOL (a), Delicious (b),  SM (c) and Twitter (d). Similar to
figure 4, the solid lines are for eye guidance.}
\end{center}
\end{figure}

The inter-event time distributions based on absolute clock are
shown in Fig. 6. For AOL and Delicious, exclusive of slightly
drooping heads, their distributions can be well approximated by
power laws. The distributions for SM and Twitter are more
complicated, with only the middle parts following power laws. The
whole distributions cannot be accurately fitted by power laws, and
the solid lines are only for eye guidance. In fact, we are not
interested in whether these distributions are power-law, but we
have noticed that the distributions typically span over six orders
of magnitude, which is more than enough to demonstrate the
burstiness of temporal activities.

Compared with Fig. 3(c), 3(d) and Fig. 2, the observed broad
inter-event time distribution may mainly result from the
heterogeneity of activity shown in Fig. 5. If so, the distribution
$p(\tau)$ should be narrowed when using the relative clock. Figure
7 reports the results of $p(\tau)$ by using the relative clock. In
accordance with the fourth point and Fig. 4 obtained from the
theoretical model, every distribution has a peak near to the head
(SM and Twitter are more remarkable). However, different from Fig.
3(c) and 3(d), the heavy tails in $p(\tau)$ cannot be weakened or
eliminated by using the relative clock. These results strongly
suggest that the observed heavy-tailed nature cannot be simply
explained by the activity heterogeneity or seasonality.

\section{Discussion}

In this Letter, we proposed a new timing method based on the
so-called relative clock, where the time interval between two
consecutive events of an individual is quantified by the number of
other individuals' events appeared during this interval. This method
is expected to be able to eliminate the effects of heterogeneity of
global activity on the inter-event time distribution. The simulation
results on the theoretical model have demonstrated the effectiveness
of our method, and by comparing the performances of simulations with
experiments, we conclude that the observed heavy-tailed nature in
human online temporal activities could not be well explained by
Poissonian agents with activity heterogeneity, no matter whether the
seasonality gets embodied in the activity time series.

Human behavior is one of the most complex and complicated things,
driven by countless unknown factors. Therefore, given a certain
statistical feature, to distinguish the effects from different
factors is very significant. Our method could successfully filter
out the effects caused by activity heterogeneity, yet it is not
omnipotent. For example, our method may lead to bias when there
exists strong trend of global activity\footnote{In online systems,
usually as the number of users increases, the global activity will
also increase, and a certain length of absolute time interval will
thus become larger and larger on relative clock.}, and thus in
these cases, detrend algorithms \cite{Peng1995,Hu2001,Chen2002}
have to be associated with our method. In addition, the effects of
heterogeneity among individuals (i.e., different individuals act
with different rates) could not be filtered out by our method,
which is also a known candidate that may contribute to the
heavy-tailed inter-event distribution \cite{Hidalgo2006}. The
rescaling method \cite{Radicchi2009,Radicchi2008} according to the
average inter-event time may be helpful in judging whether the
active and inactive individuals act with essentially different
patterns\footnote{Radicchi \cite{Radicchi2009} suggested that the
active and inactive users are of the same activity patterns since
the inter-event time distributions of high-active group and
low-active group will collapses to a single curve. Our empirical
studies show that for some data sets like FriendFeed, the
rescaling performance is not good \cite{Zhao2011b}. In fact, the
Internet users may have different behaving patterns, for example,
in online resource-sharing systems, new users tend to visit
popular things yet old users tend to dig out cool objects
\cite{Shang2010b}.}.

As a starting point of designing effective tools to distinguish the
effects of different factors on the statistical regularities of
human dynamics, the present method is simple and imperfect, yet it
may largely complement the current understanding of our behaviorial
patterns. In summary, the main contributions of this Letter are
threefold. Firstly, by using a theoretical model, we show the
heavy-tailed nature in population level may result from an exogenous
factor--the activity heterogeneity, and the timing method based on
relative clock can successfully eliminate such exogenous effects.
Secondly, extensive empirical analysis reveals the heavy-tailed
inter-event time distributions of typical online systems, and
suggests the existence of endogenous mechanisms that can not be
explained by the activity heterogeneity or seasonality versus time.
Lastly, this Letter reports many novel empirical results to the
scientific community, which could facilitate the studies on human
dynamics. Although our knowledge about human behavior increases
incessantly, it never gets sufficient. We believe this work has
added new insights and rich empirical materials into our knowledge.

\acknowledgments

This work is partially supported the National Natural Science
Foundation of China under Grant Nos. 70871082, 10975126 and
70971089, and the research grants from Research Grant Council,
University Grant Committee of the HKSAR, and Hong Kong Baptist
University.

\end{document}